\documentclass{aa}  
\usepackage{graphicx}
\usepackage{txfonts}
\usepackage{epsfig}
\begin{document}
\title{VLT Observations of NGC~1097's ''dog-leg" tidal stream\thanks{Based on observations made with ESO Telescopes at Paranal Observatory during the commissioning of FORS2.}}

\subtitle{Dwarf spheroidals and tidal streams}

\author{Pasquale Galianni\inst{1}, Ferdinando Patat\inst{2}, James L. Higdon\inst{3}, Steffen Mieske\inst{4}, Pavel Kroupa\inst{5}}


\institute {Undergraduate student, Universita del Salento, Via per Arnesano 1, 73100 Lecce, Italy.
\email{pasquale.galianni@gmail.com} 
\and ESO European Southern Observatory, Karl
Schwarzschild str. 2, D-85748 Garching bei Muenchen, Germany.
\and Department of Physics, Georgia Southern University, Statesboro, GA 30458, USA.
\and ESO European Southern Observatory, Alonso de Cordova, 3107, Vitacura, Santiago, Chile.
\and Argelander Institut f\"ur Astronomie, Auf dem H\"ugel 71, D - 53121 Bonn, Germany}

\date{Received xx, 2010; accepted xx}

\abstract
{Dwarf Spheroidal Galaxies and tidal streams.}
{We investigate the structure and stellar population of two large
stellar condensations (knots A \& B) along one of the faint optical
''jet-like" tidal streams associated with the spiral NGC~1097, with
the goal of establishing their physical association with the galaxy
and their origin.}
{We use the VLT/FORS2 to get deep $V$-band imaging and low-resolution
optical spectra of two knots along NGC~1097's northeast ''dog-leg"
tidal stream. With this data, we explore their morphology and stellar
populations.}
{Spectra were obtained for eleven sources in the field surrounding the
tidal stream. The great majority of them turned out to be background
or foreground sources, but the redshift of knot A (and perhaps of knot
B) is consistent with that of NGC~1097. Using the $V$-band image of
the ''dog-leg" tidal feature we find that the two knots match very
well the photometric scaling relations of canonical dwarf spheroidal
galaxies (dSph). From the spectral analysis we find that knot A is
mainly composed of stars near G-type, with no signs of ongoing star
formation. Comparing its spectrum with a library of high resolution
spectra of galactic GCs, we find that the stellar population of this
dSph-like object is most similar to intermediate to metal rich
galactic GCs. We find moreover, that the tidal stream shows an ''S"
shaped inflection as well as a pronounced stellar overdensity at knot
A's position. This suggests that knot A is being tidally stripped, and
populates the stellar stream with its stars.}
{We have discovered that two knots along NGC~1097's northeast tidal
stream share most of their spectral and photometric properties with
ordinary dwarf spheroidal galaxies (dSph). Moreover, we find strong
indications that the ''dog-leg" tidal stream arise from the tidal
disruption of knot A. Since it has been demonstrated that tidally
stripping dSph galaxies need to loose most of their dark matter before
starting to loose stars, we suggest that knot A is at present a
CDM-poor object. }

\titlerunning{VLT observations of NGC~1097's tidal streams}
\authorrunning{Galianni et al.}
\maketitle

\section{Introduction}

\subsection{The nature of NGC~1097's optical ''jet-like'' tidal streams}

\begin{figure*}
\resizebox{\hsize}{!}{\includegraphics{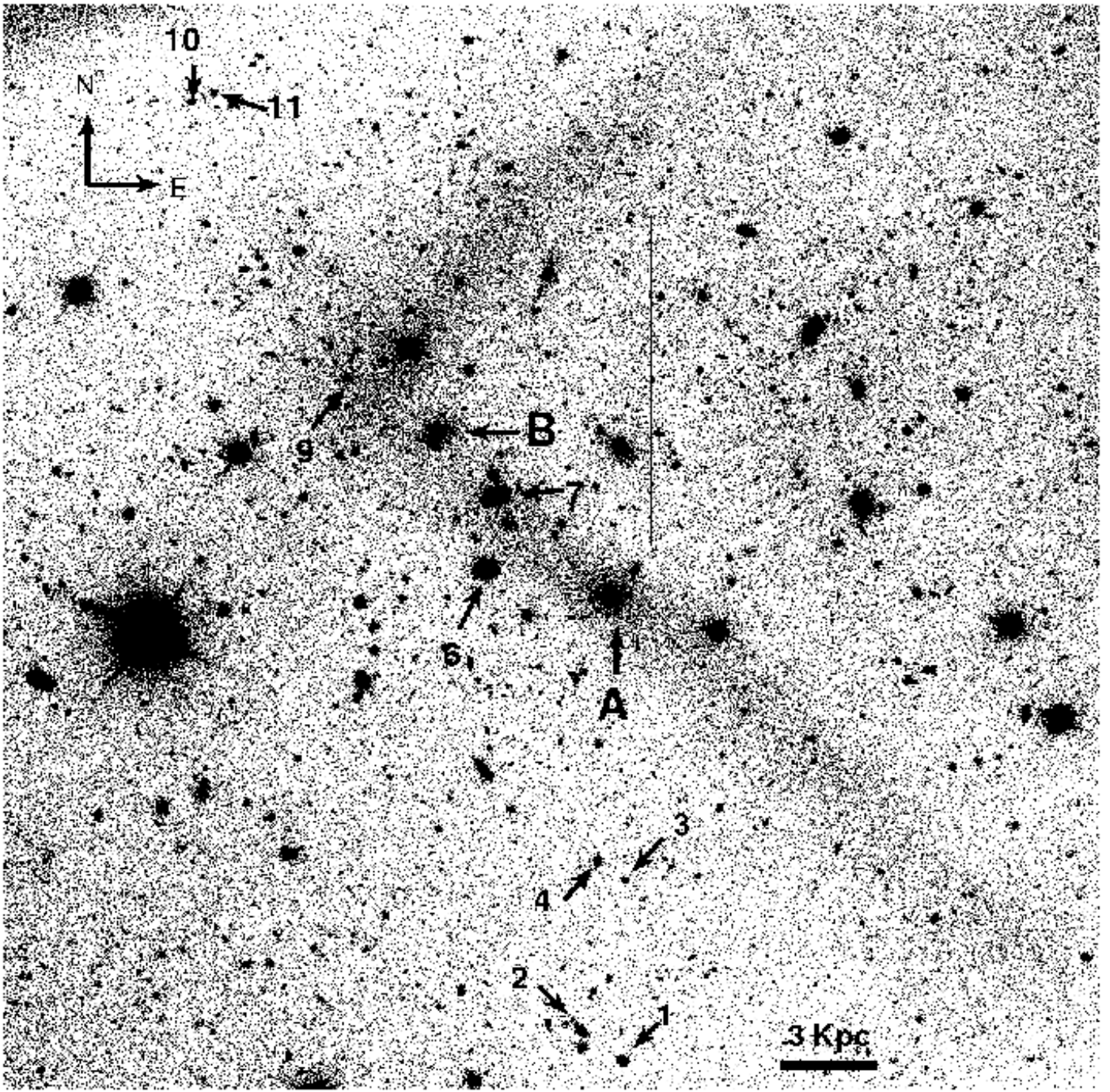}}
\caption{FORS2/UT2 400 s Bessel $V$-band image of NGC~1097's northeast optical jet and ''dog-leg".
The targets observed spectoscopically with FORS2 are labeled as in Table 1, except for knots A \&
B. The figure covers $6.8\arcmin \times 6.8\arcmin$ with North at top and East to the left. NGC~1097 lies in direction of the bottom right corner of the picture.}
\label{fig1}
\end{figure*}

		NGC~1097's network of faint optical ''jets'' have puzzled astronomers
		since their discovery in the mid-1970s (Wolstencroft \& Zealey 1975;
		Arp 1976; Lorre 1978).  These early observations established their
		blue optical colors and lack of optical emission lines. The fact that
		all four appear to radiate from NGC~1097's Seyfert 1 nucleus (see
		Fig. 1 in Lorre 1978 and Fig.\ref{fig1}) led quite naturally to
		explanations involving AGN phenomena. However, the sensitive
		upper flux limits at 1.4 GHz set by Wolstencroft et al. (1984) with the
		Very Large Array (VLA) showed that the ''jets'' optical emission did
		not arise through the synchrotron process. Their observations could
		not exclude the possibility that the ''jets'' were dominated by
		thermal Bremsstrahlung emission from a $\approx10^{6}$ K plasma (the
		high temperature is required to explain the absence of H$\alpha$
		emission set by Arp 1976). The same year, Carter, Allen \& Malin
		(1984; hereafter CAM) proposed a very different interpretation based
		on optical and near-infrared surface photometry of the two northern
		jets and the most prominent of several optical knots in the northeast
		jet first noted by Arp (1976) and Lorre (1978).  The colors of the diffuse
		light in the northern jets (e.g., $B-R = 0.6 \pm 0.3$ and $B-H =
		2.6 \pm 0.4$) are inconsistent with both thermal Bremsstrahlung and
		synchrotron emission. Instead, CAM proposed that the ''jet-like''
		features are in fact composed of {\it stars}, similar to ordinary disk
		populations ($\approx$G-type). These stars either: formed {\it in
		situ} from the cooling plasma of an ancient radio jet, were drawn out
		of NGC~1097's disk through a tidal interaction with its companion
		NGC~1097A, or represented the remains of a dwarf irregular or small
		spiral galaxy cannibalized by the much larger NGC~1097 (i.e., a minor
		merger). CAM went so far as to propose that the prominent optical knot
		near the northeast jet's abrupt right-angle bend (called the
		''dog-leg") might be what is left of the dwarf's tidally-stripped
		nucleus, given that its color ($B-R = 0.9 \pm 0.2$) is similar to
		that of late-type spiral nuclei. Wehrle et al. (1997) used VLA
		observations at 327 MHz to conclusively rule out the ''jet-like''
		features being a network of ancient radio jets, and they concluded
		that NGC~1097's jets are nothing more than a set of unusual {\it tidal
		streams} created through multiple encounters with the small elliptical
		companion NGC~1097A. Since tidal streams, and especially {\em blue}
		tidal streams, are typically rich in neutral atomic hydrogen gas (HI),
		this opened the interesting possibility of using HI kinematics to
		explore their origin and evolution.
		
		Higdon \& Wallin (2003) (hereafter HW) revived the ''minor merger'' interpretation
		for the tidal streams. Using the VLA in its most compact
		configuration, they found that all four tidal streams are extremely
		gas poor ($\Sigma_{\rm HI}$ $<$ 0.06 M$_{\odot}$pc$^{-2}$, 3
		$\sigma$). Given their blue color, they are unlike any tidal stream in
		the literature (cf. Hibbard et al. 2000; Higdon, Higdon \& Marshall 2006). 
		The total lack of HI had additional implications: the stars could not have originated from the
		HI rich disk of NGC~1097, nor could they have been formed {\em in
		situ} from a cooling radio jet without unrealistic star formation
		efficiencies. HW proposed a scenario in which the
		tidal streams were formed by multiple passes of a gas rich dwarf
		galaxy through the center of the much more massive NGC~1097. Their n-body
		simulations of such a capture produced features that strikingly
		resembled the four optical tidal streams, including the abrupt
		90$^{\circ}$ bend of the dog-leg region (see their Figures 12-14). The
		dwarf galaxy's ISM is swept out by ram pressure stripping during its
		initial pass through NGC~1097's disk, resulting in the creation of
		essentially gas free ''jet-like'' stellar streams. Whitin the HW picture,
		NGC~1097's optical tidal streams would thus represent the late stages 
		in the cannibalization of a small disk galaxy by a much larger spiral.

\subsection{Structures in NGC~1097's northeast tidal stream}

		Arp (1976) and Lorre (1978) noted the presence of several optical
		knots near the northeast tidal feature's dog-leg region (see
		Fig. \ref{fig1} and \ref{fig3}) that appeared too blue for
		background ellipticals, though with no redshifts available the possibility that
		they were background objects could not be ruled out. Wehrle et
		al. (1997) obtained 4000-7000 \AA\ spectra of the two brightest knots
		using 1-2 hour exposures with the CTIO 4 m Blanco telescope, and
		detected only weak continuum (after averaging over large wavelength
		bins) and no measurable line emission (e.g., EW$_{\rm H\alpha} < 15$
		\AA\ and EW$_{\rm [O~III]}$, EW$_{\rm H\beta} < 30$ \AA). Because of 
		detector instabilities, the quality of their spectra was not sufficient 
		to determine the nature of knots A \& B. While it had yet to be
		established that the knots were in fact part of the northeast tidal
		stream, it was clear from their apparent lack of strong emission lines
		that neither were star forming dwarf galaxies or distant AGN.  The
		existence of multiple knots are of particular interest, as they
		might represent ongoing structure formation in the tidal streams. 
		
		In this paper we report on VLT/FORS2 spectra of 
		five of the brightest optical knots in the northeast tidal stream. We
		show that the most prominent condensation, knot A, has the same
		redshift as the spiral NGC~1097, and argue that it is physically
		associated with the tidal stream. A second condensation, knot B, is
		also plausibly associated with the tidal feature.$\footnotemark[1]
		$\footnotetext[1]{knots A \& B referred to in this paper correspond to
		the two optical knots discussed in Wehrle et al. (1997). Our knot A is
		also the ''bright condensation'' in jet R1 discussed by CAM (see their
		Section 3 and Table 2).} 
				
		The VLT observations are described in Section 2. In Section 3 we
		present the photometric and spectroscopic measurements of knots A \&
		B, and discuss these findings and their implications in Section
		4. Finally, we summarize our results in Section 5. Throughout this
		paper we have adopted the standard WMAP cosmology ($H_0$=$72~\pm~5$ km
		s$^{-1}$ Mpc$^{-1}$; Spergel et al. 2003), which for NGC~1097's redshift ($z=0.0042 \pm
		0.0001$, e.g., Koribalski et al. 2004; Mathewson \& Ford 1996) results
		in a luminosity distance $D_{\rm L}$ of 17 $\pm~2$ Mpc and a linear scale of
		84 pc/$''$.

\section{VLT observations and data reduction}

	The observations were made at ESO-Paranal by
	G. Rupprecht and H. Arp on several runs: 7 October 2000 (ID 101443), 
	17 November 2000 (ID 103791) and 4 December 2000 (ID 103790).
	The data were kindly provided by Arp and Rupprecht.
	Optical spectra were obtained with the FORS2 imaging-spectrograph
	(\cite{app98}), situated at the Cassegrain focus of the 8.2 m VLT Kueyen
	(UT2). The detector was a 2048x2048 TK2048EB4-1 thinned, backside illuminated CCD
	 and the integration time of our spectra was of 30 minutes.
	The standard resolution collimator was used, providing an angular 
	scale of 0.2 $''$/pix and a field of view of 6\arcmin.8x6\arcmin.8. 
	Grism GRIS\_150I+27 was used, which provides a linear dispersion of 230 \AA/mm 
	and $\lambda/\Delta\lambda = 260\pm1.5$ (if coupled with a 1\arcsec ~ slit).
	Spectra covering the full 3300-1000 \AA ~wavelength range
	were obtained in two stages: red spectra (6000-10,000 \AA) using the
	OG590+32 filter as an order blocker, and blue spectra (3300-6600 \AA)
	with no filter. 
		
	The observations were carried out in multi-object (MXU) mode, with
	twelve slits of varying widths (1$''$ to 2.5$''$) placed on the
	sky. One large slit was situated across the northeast tidal stream, one slit
	each was placed on knots A \& B, six on field objects, and three on
	empty fields to measure sky emission. Data reduction was routine, 
	and standard procedures in IRAF\footnotetext[2]{IRAF is distributed by the National 
	Optical Astronomy Observatories, which are operated by the Association of 
	Universities for Research in Astronomy, under contract with the National Science Foundation.}
  were used to extract (\textit{apall}),
	calibrate (\textit{identify}, \textit{calibrate}), and join (\textit{scombine}) 
	the red and blue spectrum for each slit. 
	See Table \ref{table1} for the coordinates,
	photometry and a short description of the observed objects. 
	
\begin{figure*}[h!]
\begin{center}
  \epsfig{figure=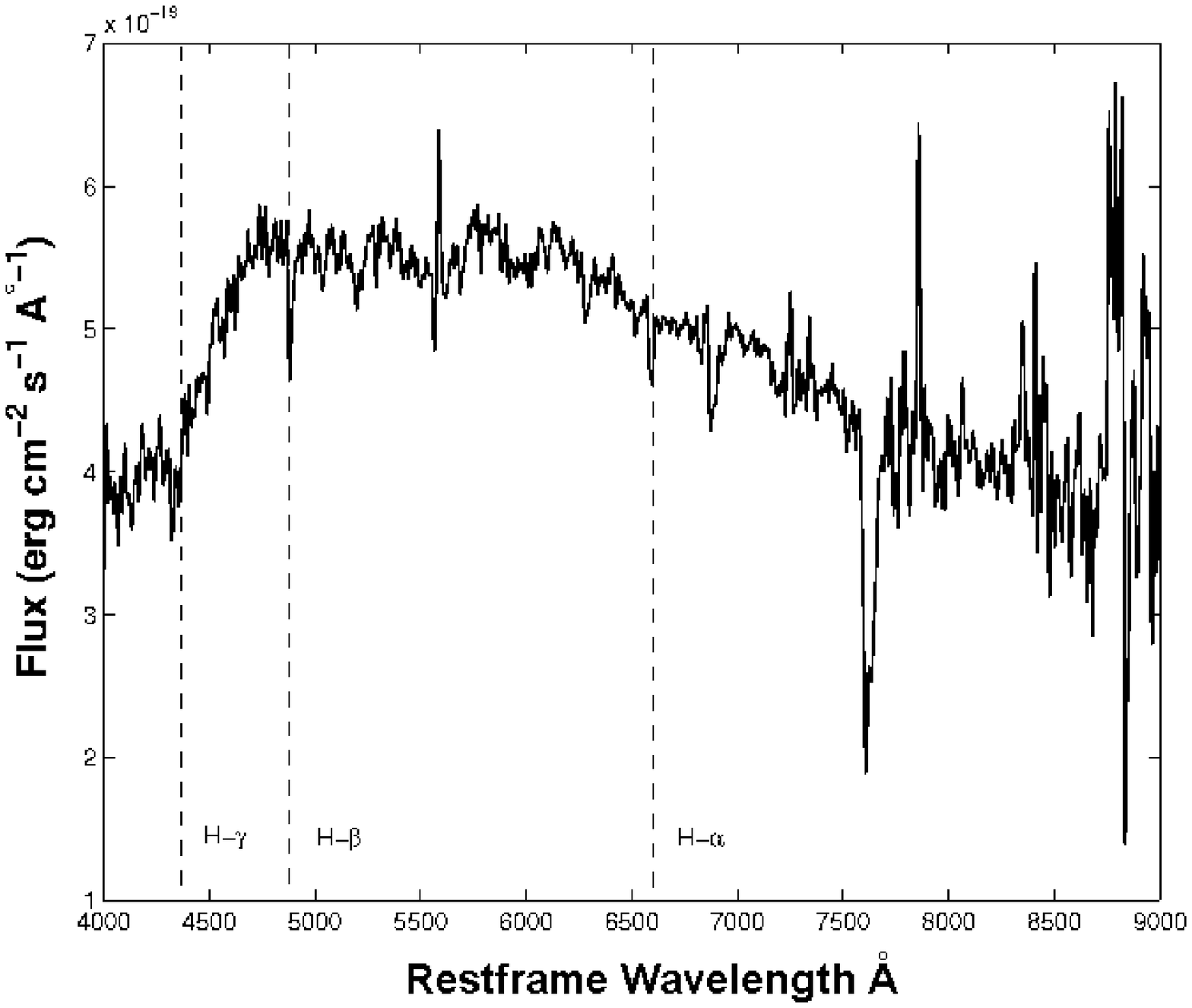,width=9cm}
  \epsfig{figure=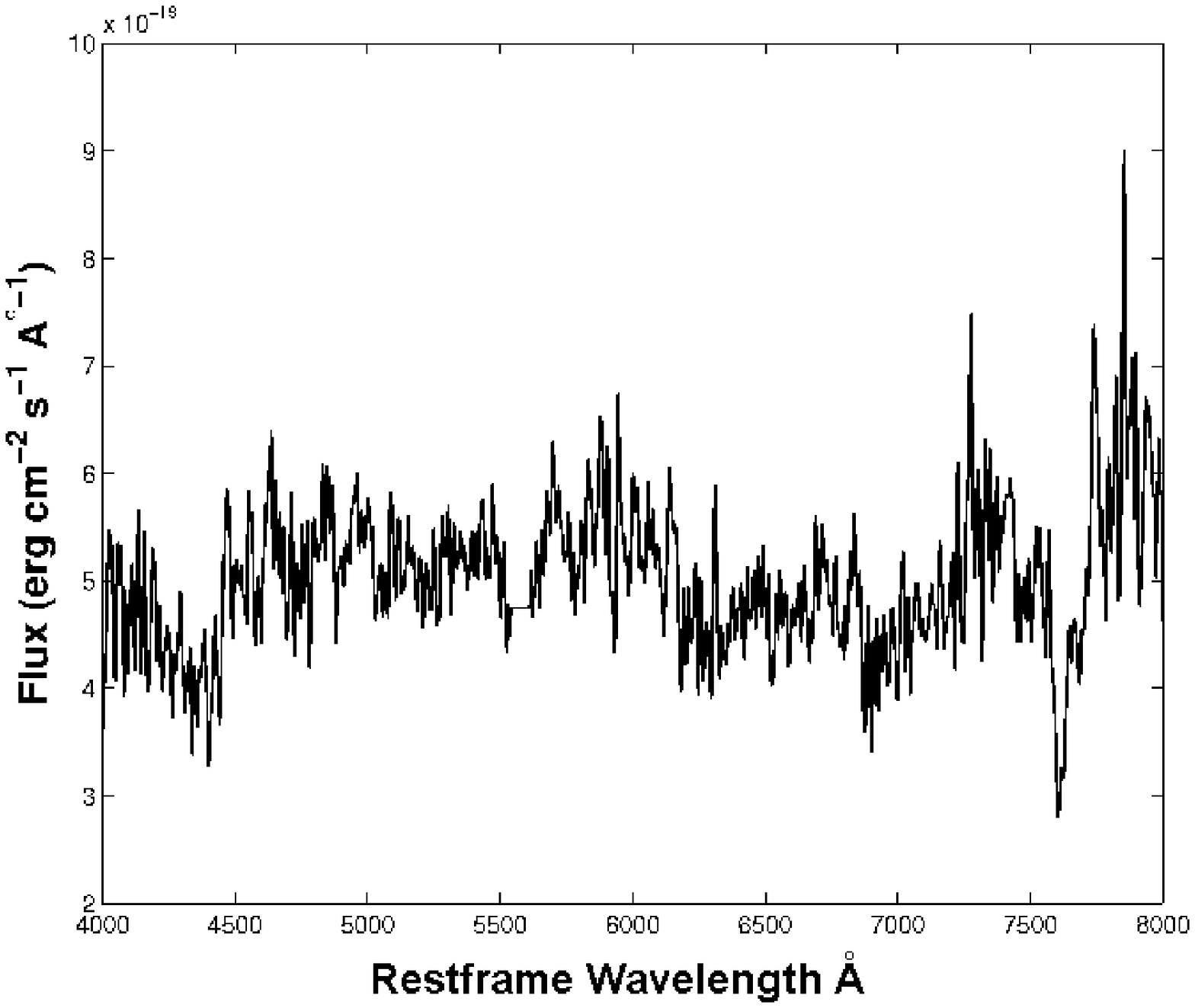,width=9cm}
  \epsfig{figure=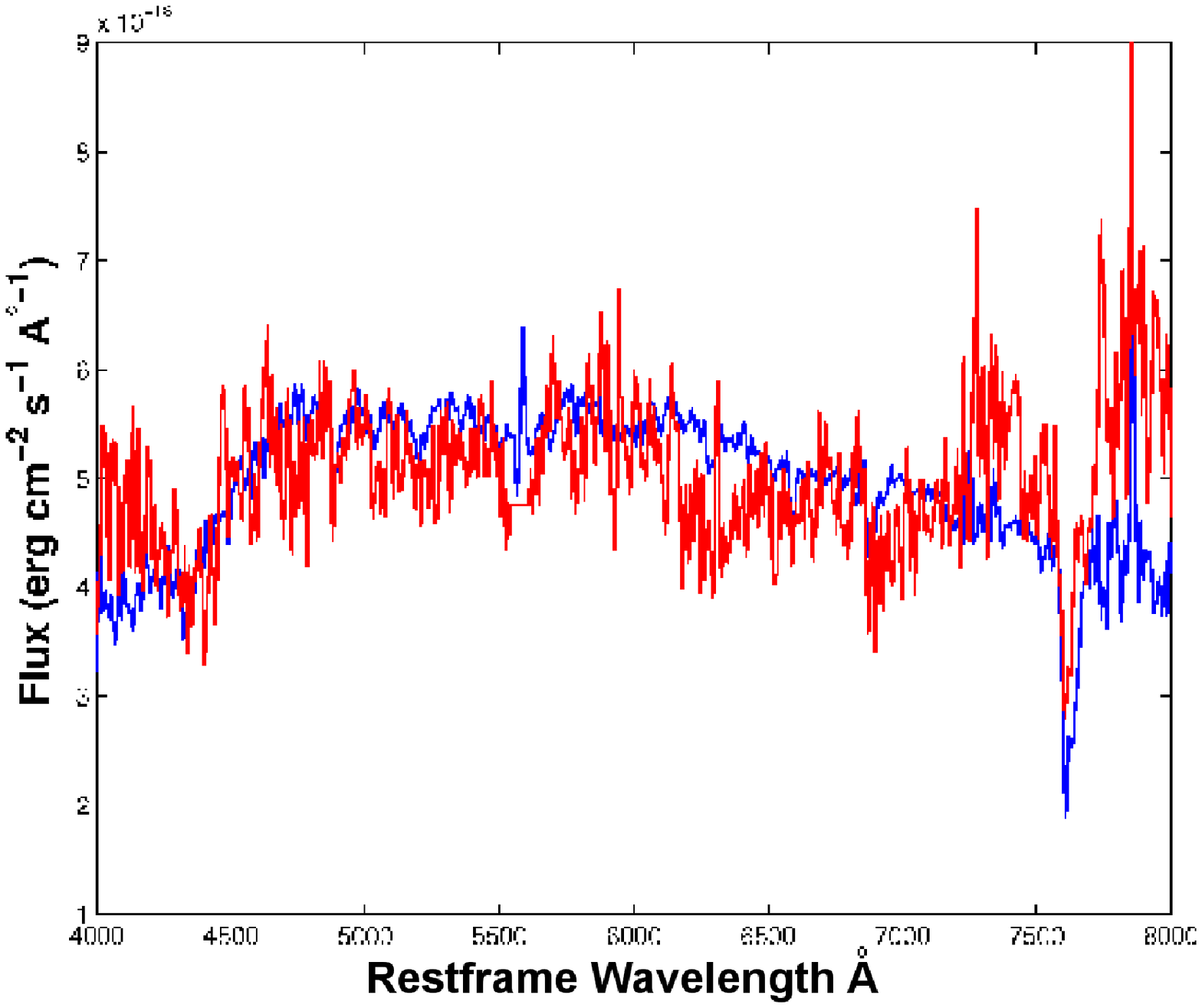,width=12cm}
  \caption{\textit{Up Left:} FORS2 spectrum of knot A. \textit{Up Right:} FORS2 spectrum of knot B.
  \textit{Bottom:} Comparison of knot A's spectrum (blue line) vs knot B (red line).}
\label{fig2}
\end{center}
\end{figure*}

\begin{figure*}[h!]
\begin{center}
  \epsfig{figure=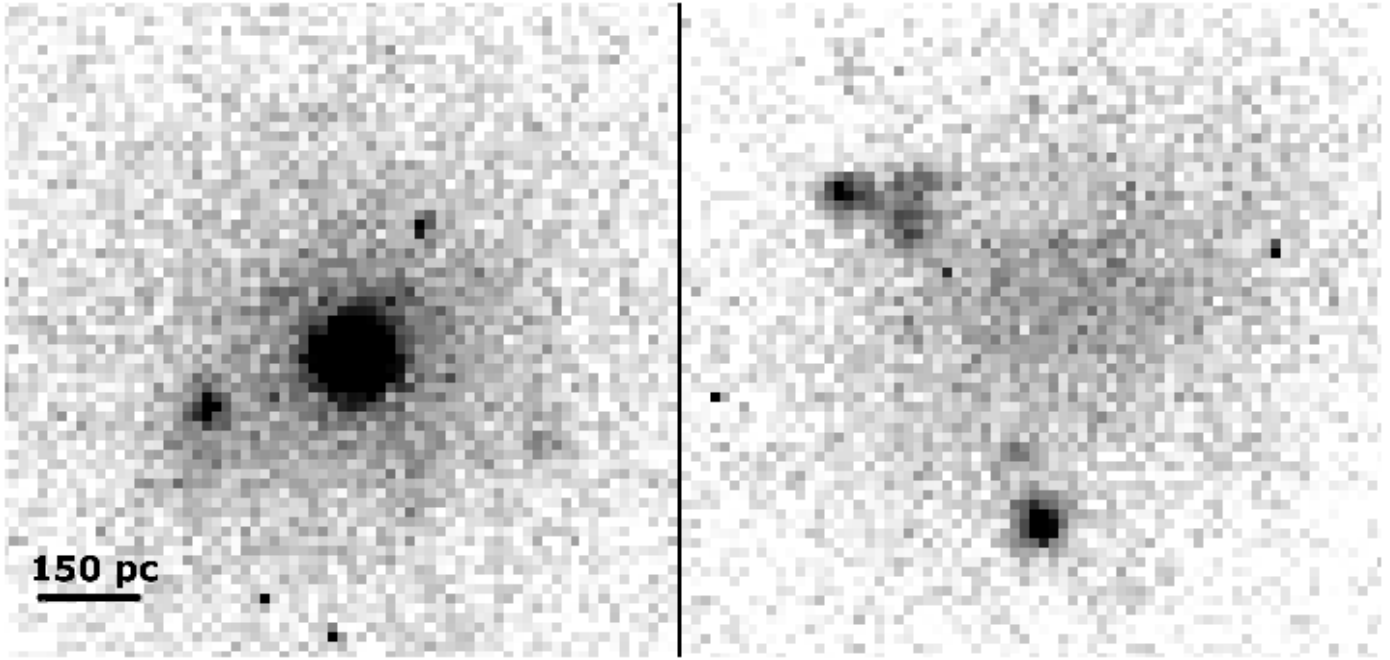}
  \caption{Closeups ($13\arcsec \times 13\arcsec$) from the Bessel $V$-band image of NGC~1097's 
  dog-leg region (Fig. 1) showing (left) knot A and (right) knot B. Identical linear stretches
  are used.}
\label{fig3}
\end{center}
\end{figure*}

	The spectrum of the northeast tidal stream was too faint to be 
	succesfully extracted with \textit{apall}. Despite using the largest 
	available slit and the lowest dispersion grism available, the tidal stream proved too faint 
	for useful spectroscopy in 30 minutes of integration. We
	obtained, however, well exposed spectra for nine other objects, including
	knot A, and a less exposed but still useful spectrum of knot B (see Fig. \ref{fig2}).
	FORS2 was also used to obtain a 400 s Bessel $V$-band exposure
	centered on the northeast dog-leg tidal stream (see Fig. \ref{fig1}). The
	night's seeing (FWHM of unsaturated stars measured on the image) 
	was $~\approx1\arcsec$. 

\section{Results}

\subsection{Spectroscopy of the condensations in the ''dog-leg" tidal stream}
	
		The five optical condensations in the northeast tidal stream that were observed
		spectroscopically with FORS2 are indicated in Fig. 1 and listed in
		Table 1.  Of these, one is a foreground star (Object 9) and two are background galaxies 
		(Objects 6 \& 7), with $z \approx 0.1-0.3$. We will not discuss these sources further. 
				
		A high quality spectrum was extracted for knot A, and is shown in
		Fig. 2. The spectrum is notable for a total lack of emission lines
		ordinarily found in star forming systems like [O~III] $\lambda$
		4959, 5007 \AA, H$\beta$ or H$\alpha$, in agreement with Arp (1976) and
		Wehrle et al. (1997). However, these new observations set more
		stringent limits on H$\alpha$ emission, with EW$_{\rm H\alpha} < 5$
		\AA\ . Several narrow hydrogen Balmer absorption lines (H$\gamma$,
		H$\beta$, and H$\alpha$) are clearly detected, with equivalent widths
		(measured with IRAF's \textit{splot} tool) of 2.9, 3.8 and 2.3 \AA\, 
		respectively (see Table 2).
		There is also evidence for a weak G-band ($\lambda$ 4303 \AA)
		absorption. The most prominent feature in Fig. 2 is a strong break in the 
		continuum level at $\approx$4400 \AA. 
		
		We derive synthetic optical colors for knot A using the spectrum in Fig. 2
  	by numerically integrating over the Johnson-Cousins UBVRI passbands, and find
  	$B-V = 0.8 \pm 0.1$, $B-R = 1.3 \pm 0.1$, $V-R = 0.5 \pm 0.1$ and $V-I = 0.9 \pm 0.1$.
  	Note that these colors are somewhat redder than the colors of the local diffuse jet
  	emission as measured by CAM ($B-R = 0.9 \pm 0.2$). The 1.8 $\sigma$~deviation between our and CAM's 
  	measures is not surprising considering the inherent difficulty in measuring the colors of such a low brightness
  	feature, and the fact that CAM performed those measures more than 20 yrs ago using normal photographic plates.
						
		From the measured wavelengths of absorption lines in knot A's spectrum
		(see Table \ref{table2}) we derive a redshift of $z=0.0043 \pm
		0.0001$. This is within $\Delta z = 0.0001$ of NGC~1097's redshift
		derived using HI and optical emission lines, and shows that knot A is
		indeed physically associated with the barred spiral galaxy. 
		
		As shown in Fig. 2 (upper-right panel) the two features in the spectrum of knot B, being possibly 
		significant above the noise are the break in the continuum level at 4400 \AA~ and the H$\beta$ line
		at 4882 \AA. As is shown in Fig. 2 (bottom panel) the overall continuum shape and the position of these two
		features match fairly well those of knot A, indicating that their redshifts could be similar. 
		Under the assumption that knot B is at the same redshift distance of knot A, we find moreover that knot B 
		agree very well with the same photometric scaling relations of knot A. This makes it unlikely 
		that knot B is a galaxy with a different absolute magnitude than knot A, placed at a different 
		distance from NGC 1097.	We will therefore assume throughout the rest of the article that knot B is 
		physically associated with NGC 1097.

\subsection {Surface Photometry and morphology of knots A \& B}

		Both knots A \& B in the NE tidal stream are easily seen in the FORS2 $V$-band
		image shown in Fig. \ref{fig1}. We are able to extract new details
		concerning their morphologies: Knot A shows considerable spherical
		symmetry, with a bright and compact core and a halo that extends for
		the full width of the stream ($\approx15\arcsec$), while knot B is more
		diffuse and lacks a central core (see Fig. \ref{fig3}).
		
		Averaged $V$-band surface brightness profiles (SBPs) for knots A \& B were extracted 
		using the IRAF/STSDAS task {\em ellipse} and are shown in Fig. 4.
		A small number of faint and unresolved objects surrounding the nuclei of both knots
		are apparent in Fig. 3. These were excluded from the SBP extraction using a pixel mask.
		We fit the SBPs of both knots using the $R^{1/n}$ parameterization of S\'ersic (1968), 
		which can be written:
		\begin{equation}
		\mu (R) = \mu_{e} \exp \left\{-b_{n}\left[\left(R \over R_e\right)^{1/n}-1\right]\right\},
		\end{equation}
		where $R$ is the projected radial distance from the center of the galaxy.
		This representation is widely used and has the advantage 
		of precisely describing a variety of SBPs, including pure exponential and
		de Vaucouleurs $R^{1/4}$ laws, i.e., those of both dwarf and luminous elliptical 
		galaxies (Faber \& Lin 1983; de Vaucouleurs 1948, 1959). 
		The free parameters of this model are $R_{e}$, $\mu_{e}$ and $n$, where n describes 
		the shape of the distribution. The coefficient $b_n$ depends on $n$ and 
		can be chosen in such a way that $R_{e}$ and $\mu_{e}$ coincides respectively with 
		the half light radius of the object (i.e. the radius enclosing half of the total flux) and
		the surface brightness at that radius. As shown by Capaccioli (1987) for $1<n<10$ we have 
		$b_{n} \approx 1.9992n - 0.3271$. Since for our objects $n<1$ a question is wheter this 
		approximation is appropriate. As shown by Graham (2001) the difference between the exact 
		value of $b_{n}(n)$ and Capaccioli's approximation is about $b_n(exact) - b_n(approx) =0.03$ 
		for $n=0.4$ (knot B).
		Considering the uncertanities of our fits (RMSE$\approx0.1$) we have neglected such small
		difference.  
					
		The SBPs for knots A \& B are shown in Fig. 4, along with the
		fitted S\'ersic model and residuals. The latter are small with no systematics
		and indicate good fits in both objects. Derived values of $n$, $R_{e}$,
		and $\mu_{e}$ and their uncertainties are included in Table 3. Knot A
		possesses a bright and marginally resolved core that is not modeled by a S\'ersic 
		profile. For this reason we restricted the fitting region to exclude the inner 
		$1.6\arcsec$, which in figure 4 can be seen to correspond to the radius at which the
		fitted S\'ersic profile begins to depart from the data points, and agrees very well
		with the optical size of the bright core (Fig. 3). Knots A \& B have fitted S\'ersic
		exponents $n_{A} = 0.6 \pm 0.1$ and $n_{B} = 0.4 \pm 0.1$, and central $V$-band surface
		brightnesses (i.e., extrapolating the fitted S\'ersic profile to $R = 0$ for knot A)
		of $\mu_{A,0} = 24.6 \pm 0.2$ mag arcsec$^{-2}$ and $\mu_{B,0} = 25.1 \pm 0.2$ mag
		arcsec$^{-2}$.

		Accurate half-light radii have been computed for both objects. For knot B, we estimate
		the half-light radius from the S\'ersic model fit to be (see above) 
		$R_{e}= 428 \pm 84$ pc, where we have included contributions from the pixel 
		scale and saturation effects in the uncertainty.  
		Since knot A's core cannot be fit by a S\'ersic profile, we estimate
		an empirical half-light radius using the {\em ellipse} output to calculate the radius
		at which the total flux drops by half. In this way we obtain $R_{e} = 336 \pm 84$ pc.
	  		
		Integrated apparent $V$-band magnitudes were determined to be $m_{V} = 19.9 \pm 0.1$ mag and
		$m_{V} = 20.4 \pm 0.1$ mag for knots A \& B respectively, which correspond to absolute
		magnitudes of $M_V = -11.2 \pm 0.1$ and $M_V = -10.8 \pm 0.1$ at the adopted distance
		of NGC~1097. Assuming $M_{\odot, V} = 4.82$ (Bell \& de Jong 2001), these translate into
		$V$-band luminosities of $L_{V} = 2.6 \times 10^{6}$ L$_{\odot,V}$ for knot A
		and $L_{V} = 1.6 \times 10^{6}$ L$_{\odot,V}$ for knot B. We also estimate $m_V$, $M_V$,
		and $L_V$ for knot A's core using a circular aperture of radius $3 \times R_{core}$, and find 
		$m_V = 21.4 \pm 0.1$ mag, $M_V = -9.8 \pm 0.1$ and $L_V = 6.9 \times 10^{5}$ L$_{\odot,V}$.
		
		Dwarf galaxies subject to ongoing tidal perturbations may show surface brightness
		``breaks'' or other irregularities in their outer isophotes Pe\~narrubia et al. (2009).
		We are however unable to detect such irregularities in our surface brightness profiles (see Fig. 4).
		
		This however does not necessarly imply that the knots are not being tidally stripped, since
		as shown by (Pe\~narrubia et al. 2009) such ''bumps" in the isophotes are essentially transient
		features that quickly drift in the outer region of the SBPs, where the surface brightness 
		is too low to give any valuable information. Their eventual detection is therefore related to the
		time of the observation, and strongly depends on the orbital parameters of the tidally-disrupting object.
		Moreover, it is likely that the spatial resolution of our SBPs is simply insufficient to reveal these ''bumps": 
		our objects have in fact a very small angular extension if compared to Local Group (LG) dwarfs. It is however difficult to 
		estimate the expected amplitude of these irregularities without knowing the orbit of the objects and their internal
		kinematics.

\subsection {Stellar population}
		
		The resolution of knot A's spectrum ($\delta\lambda = 19$ \AA\ at
	  5000 \AA) is too low to accurately estimate metal abundances from
	  spectral line indices. An estimation of the metallicity is nonetheless
	  important to ascertain the nature of the objects.
	  
	  In order to get more informations about knot A's stellar population, we have 
	  cross-correlated its spectrum with that of 40 galactic GCs from the library presented
	  in Schiavon et al. (2005) which covers a wide range of metallicities from -2 dex to solar abundance.
	  We degraded the spectral resolution of the 40 galactic GCs to match that of knot A's spectrum, and
	  used the IRAF task \textit{fxcor} for cross-correlation. 
	  
	  With this method we find a linear relation between the cross correlation amplitudes of knot A 
	  (plus two test GCs) and the [Fe/H] ratios of the library's GCs (see Fig. 5). This is reasonable considering that for 
	  evolved stellar populations, the [Fe/H] ratio should play an important role in determining spectral differences. 
	  This implies that the [Fe/H] ratio of knot A can be estimated --at least qualitatively-- using this cross correlation 
	  technique. 
	  
	  The results of the left panel of Fig.5 imply an [Fe/H] ratio $>$-1.0 dex for knot A. The analogous plots
	  in the middle and right panel for a metal-poor and a metal-rich GC confirm the validity of this approach, since the slope
	  of the cross-correlation amplitude vs. [Fe/H] is, as expected, negative for the metal-poor and positive for the metal-rich
	  GCs.
	  	  
    Fig. 6 confirms the results of Fig. 5 by comparing knot A's spectrum with the two GCs with highest and lowest
    cross-correlation amplitudes. The spectra match very well for the GC NGC 6388 which is of intermediate metallicity
		([Fe/H]=-0.7 dex), and show a clear discrepancy for NGC 1904 which has a lower metallicity ([Fe/H]=-1.5 dex).
		Since NGC 6388 has an integrated spectral type of G2, while NGC 1904 has type F4/5 (Harris et al. 1996), we can state 
		that the light emitted by knot A's is probably dominated by G-type stars, in agreement with CAM.
				
		While it is true that this method does not precisely determine the [Fe/H] ratio, the results shown
    in Fig. 5 \& 6 indicate (at least qualitatively) that knot A's metal abundances are higher than LG
    dwarf spheroidals of similar luminosity (e.g., [Fe/H]= -1.5 dex; Mateo 1998). It has been shown
		however, that dSph galaxies belonging to different clusters of galaxies may show sensible 
		differences in their metallicity-luminosity relation, if compared with LG dwarves (Lianou, Grebel \& Koch 2010).
				  
\section{Discussion}

\subsection{Are the knots dwarf spheroidal galaxies?}

	We have shown that knot A's optical spectrum and total luminosity
	matches well that of intermediate to metal rich and massive GCs like 47 Tucanae and Mayall I.
	However, the size of knot A ($R_{e}~=~337~\pm~84$ pc) is abundantly 
	beyond those of ordinary GCs, the vast majority of which possess 
	$R_{e} < 10$ pc (cf. Mackey \& van den Bergh 2005). In terms of 
	size both knots A \& B are similar to LG dSph satellite galaxies of comparable
	luminosity (\cite{mat98}).
	It has been established that dSph galaxies and GCs occupy different
	positions in a plot of half-light radius versus $M_{\rm V}$. Large GCs
	in fact obey a well defined relation, in the sense that larger GCs are
	also fainter ($log[R_{\rm 50}]~=~0.25~M_{\rm V}~+~2.5$; Mackey \& van
	den Bergh (2005), though Van den Bergh (2008) discusses shortcomings
	of this diagnostic). Knots A \& B are well above this line trend (they
	are much larger for their optical luminosity than GCs) (see Fig. 7).
	Instead, the knots agree very well with the $M_v-\mu_{0,V}$ relation for LG 
	and Hydra/Centaurus dwarves (Misgeld et al 2008, 2009).
	They found that typical dSph galaxies follow the relation (cf. Misgeld et al. 2009):
	\begin{equation}
	\mu_{0,V}*=0.57[\pm0.07]*M_V + 30.90[\pm0.87]
	\end{equation}
	Substituting $M_V=-11.2\pm0.1$ for knot A and $M_V=-10.8\pm0.1$ for
	knot B, we obtain $\mu_{0,A}*=24.6\pm0.085$ mag arcsec$^{-2}$ and $\mu_{0,B}*=24.74\pm0.085$
	mag arcsec$^{-2}$ that match very well our measured central surface brightnesses (see Table 3).
	Our S\'ersic indexes of $n_{A} = 0.6 \pm 0.1$ and $n_{B} = 0.4 \pm 0.1$ also agree very well with the typical 
	values for Hydra/Centaurus dwarfs with a $M_v\approx-11$ that is (cf. Misgeld et al. 2009) $n = 0.5 \pm 0.1$. 
	
  In terms of stellar population (old GC-like stellar population with no signs of ongoing SF and a peculiar lack of
  HI), stellar mass ($M_{*}\approx6~\times~10^6 M_{\sun}$ for knot A and $M_{*}\approx4~\times~10^6 M_{\sun}$ for
  knot B), central surface brightness and S\'ersic index (see above), $M_v$ vs $R_{50}$ (see Fig. 7) our knots closely resemble
  ordinary dSph galaxies, as defined in Grebel, Gallagher \& Harbeck (2003) and Mateo (1998). 
	
\subsection{Structure and composition of the stellar stream}
  
  The stellar stream itself was too faint for quantitative spectroscopy. 
  From our sky-substracted $V$-band exposure however (see Fig. 1), we could 
	measure the mean surface brightness of the stream (measured over ten different apertures along the ''dog-leg"). The 
	value that we obtained is $\mu_{V}=26.5 \pm 0.2$ mag/arcsec$^2$.
	After measuring the mean SB of the stream using the standard IRAF tools, authors PG and SM independently measured the size of
	the stream, by subdividing it in small rectangular apertures. The value found is $8400 \pm 100$ arcsec$^2$.
	From the measurement of the stream's mean surface brightness and its area, we have calculated its integrated V magnitude to 
	be $m_{V}=16.6 \pm 0.2$ mag. At the distance of NGC 1097 this corresponds to an absolute magnitude of $M_{V} = -14.5\pm0.35$
	mag.
	
	Useful hints about the composition of the tidal stream can be found in former studies 
	(CAM, HW, \cite{weh97}). Using multiband photometry of the tidal stream (obtained in a region 
	slightly south-western than knot A) CAM suggested that the SED of the ''dog-leg" feature, is compatible with G-type stars.
	They also found that the colors of the stream and knot A are compatible within 
	the errors ($B-R_{str.}=0.55\pm0.36$ , $B-R_{A}=0.9\pm0.18$).
	
	This last point agrees with the conclusions of Wehrle at al. (1997). 
	In their paper they measured the B/V count ratio longitudinally and transversally 
	along the tidal stream (see their Fig. 8), and concluded: 
	''The color [along the tidal stream] is constant within the errors, including both prominent
	condensations".	
  These two studies came to the conclusion that the stellar stream is composed of stars near G-type 
  and that the stream and the knots have the same color. With our FORS2 spectra, we independently showed that also knot A is
  predominantly composed of stars near G-type. This suggests that the knots and the tidal stream are both composed of the same
  stellar material.
  
  A morphological analysis of the tidal stream indicates that knot A is currently being tidally stripped,
  populating the ''dog-leg" tidal stream with stars. As shown in Fig. 8 (left) the tidal stream show a slight but significant
  ''S" shaped inflection coincident with the position of knot A. In Fig. 8 (right) we show moreover the elliptical overdensity
  of stars in correspondence of knot A. These morphological features are typical for tidally disrupting systems (\cite{for03},
  \cite{mar08}, 2010).

  If knot A is the only progenitor of the stellar stream, before the encounter with NGC 1097, knot A should have been a dwarf
  galaxy of at least $M_{V} = -14.5 \pm 0.35$ mag. This means that knot A has lost at least the 95\% of its stars during the
  encounter with NGC 1097. This is in agreement with the n-body simulations performed by HW that shows how
  such a peculiar structure like the ''dog-leg" tidal stream can have arises from the tidal disruption of a low mass galaxy.
   	
\subsection{How did the knots form?}
	
  The alignment of knots A \& B with the ''dog-leg" tidal stream suggests that these two objects
  are probably correlated in phase-space. Such a perfect alignment along the same stream would be in fact 
  very unlikely for independently infalling CDM-Subhalos.  
  
  A possible explaination to the phase-space correlation problem of MW satellites (\cite{kro05}; \cite{mk09}), has been
  proposed in terms of a ''group infall" of sub-halos (Li \& Helmi 2008; D'Onghia \& Lake 2008) or suggesting that dwarf
  galaxies form along dark matter filaments (Ricotti et al. 2008). It is however still unclear if these mechanisms can
  efficiently explain the observed distribution of satellite galaxies around the MW and Andromeda (for recent criticism see:
  \cite{mz09}; \cite{mk09}).
  
  The alignment of knots A \& B with the stream is instead reminescent of the situation in the Milky Way, where the
  disk-of-satellites is approximately aligned with the Magellanic Stream (\cite{mz09}). 
  This may suggest that the satellite galaxies of NGC 1097 may also be interpreted as being old tidal dwarf galaxies (Zwicky
  1956; Lynden-Bell 1983; Okazaki \& Taniguchi 2000). 
  
  However, a definitive interpretation awaits further study, in particular, we need to examine the knot's
  internal kinematics, and whether NGC 1097 may have further dSph satellite galaxies. The deep implications for fundamental
  physics of  objects  such as knots A \& B being tidal dwarf galaxies are discussed in Kroupa et al. (2010).
	
\section{Conclusions}

	We have shown that the two optical ''knots'' along NGC~1097's tidal
	stream share most of their observable properties with ordinary dwarf spheroidal galaxies (dSphs). 
	From the measured redshifts we show that knot A (and very likely knot B) are associated
	with the tidal stream. The spectral light distribution of these dSphs is
	most consistent with that of intermediate to metal-rich Galactic GCs (see Fig.s 6, 7).
	
	These new observations set more stringent limits for H$\alpha$ emission of 
	the tidal stream, with EW$_{\rm H\alpha} < 5$ \AA.
	 
	Our new observations, togheter with the results from former studies (\cite{car84}, \cite{weh97}, \cite{hig03})
	indicates that knot A is composed of the same stellar material as the tidal stream. Moreover a morphological analisys
	of the tidal stream reveals typical signs of ongoing tidal stripping (see Fig. 8). Based on this evidences we conclude
	that very likely the stellar stream is populated by stars trailed out from knot A.
	
	The presence of an ongoing tidal stripping is incompatible with knot A being surrounded at present by a massive CDM halo
	(\cite{pen08}, \cite{pen09}).
	
\begin{acknowledgements}
	
	We want to thank Halton C. Arp (MPA Garching) and Gero Rupprecht (ESO
	Garching), E.M. Burbidge and V. Jukkarinen (CASS San Diego) for their
	useful comments. We wish to extend a special word of thanks to Martino
	Romaniello (ESO Garching), Dieter Horns and Andrea Santangelo (IAAT Tuebingen)
	for their help during the firsts development stages of this paper.
	This work is based on  observations made with ESO Telescopes at Paranal 
	Observatory during the commissioning of FORS2.
	
\end{acknowledgements}

\begin{table*}
\caption{Magnitudes of the observed Objects: The labels fs, bg, c indicates
respectively: 'foreground star', 'backgroung galaxy', 'condensation'. The Typical RMS errors on the magnitudes are 0.1 mag.}
\label{table1} 
\centering 
\begin{tabular}{c c c c c c} 
\hline\hline 
Obj. ID & RA & DEC $(J2000)$ & $m_{v}$ (mag) & Type \\ 
\hline 
1  & 02:47:00 & -30:13:04 & 21.3 & fs\\ 
2  & 02:47:01 & -30:12:59 & 19.9 & bg\\
3  & 02:46:59 & -30:11:55 & 21.4 & fs\\
4  & 02:47:01 & -30:11:48 & 21.2 & bg\\
5(knot A)  & 02:47:00 &  -30:10:08 & 19.9 & c\\
6  & 02:47:04 & -30:09:30 &  20.0 & bg\\
7  & 02:47:04 & -30:09:22 &  21.5 & bg\\
8(knot. B)  & 02:47:05 & -30:09:07 &  20.4 & c\\
9  & 02:47:07 & -30:08:38 & 22.1 & fs\\
10  & 02:47:12 & -30:06:58  & 23.4 & bg\\
11  & 02:47:11 & -30:06:57  & 23.6 & bg\\
\hline 
\end{tabular}
\end{table*}

\begin{table*}
\caption{Absorption redshift of knot A: $\lambda_{meas}$ is the measured wavelength, 
$\lambda_{0}$ is the restframe wavelength, $z$ is the measured redshift and $\Delta z$ is 
the redshift difference with NGC~1097 (z= 0.0042 NED). } 
\label{table2} 
\centering 
\begin{tabular}{c c c c c c} 
\hline\hline 
Line & $\lambda_{meas}$ & $\lambda_{0}$ & z & $\Delta z$ & EW (\AA)\\ 
\hline 
$H_{\alpha}$  & 6592.0 & 6562.8 & 0.0044 & 0.0002 & 2.9\\ 
$H_{\beta}$  & 4882.1 & 4861.3 & 0.0043 & 0.0001 & 3.8\\
$H_{\gamma}$  & 4359.1 & 4340.5 & 0.0043 & 0.0001 & 2.3\\
\hline 
\end{tabular}
\end{table*}

\begin{table*}
\caption{Summary of measured and estimated (E) parameters for knots A \& B} 
\label{table3} 
\centering 
\begin{tabular}{c c c} %
\hline\hline
Parameter name & knot A & knot B  \\
\hline 
Redshift & $z=0.0042\pm0.0001$ &  $=$ \\
Heliocentric distance & $D=17\pm2$ Mpc & $=$\\
Integrated V magnitude & $m_{v}=19.9\pm0.1$ mag & $m_{v}=20.4\pm0.1$ mag\\
Nucleus V magnitude & $m_{v_{nuc}}=21.4\pm0.1$ mag & --\\
Absolute magnitude & $M_{V}=-11.2\pm0.1$ mag & $M_{V}=-10.8\pm0.1$ mag\\
Half light radius & $R_{e}=336\pm84$ pc & $R_{e}=482\pm84$ pc\\
Central surface brightness & $\mu_{A,0} = 24.6\pm0.2$ mag arcsec$^{-2}$ & $\mu_{B,0} = 25.1\pm0.2$ mag arcsec$^{-2}$\\   
S\'ersic index & $n=0.6\pm0.1$ & $n=0.4\pm0.1$\\
Predominant stellar type & G & =\\
Stellar mass & $M_{*}\approx6~\times~10^6 M_{\sun}$ & $M_{*}\approx4~\times~10^6 M_{\sun}$\\
\hline 
\end{tabular}
\end{table*}

\begin{figure*}[h!]
\begin{center}
  \epsfig{figure=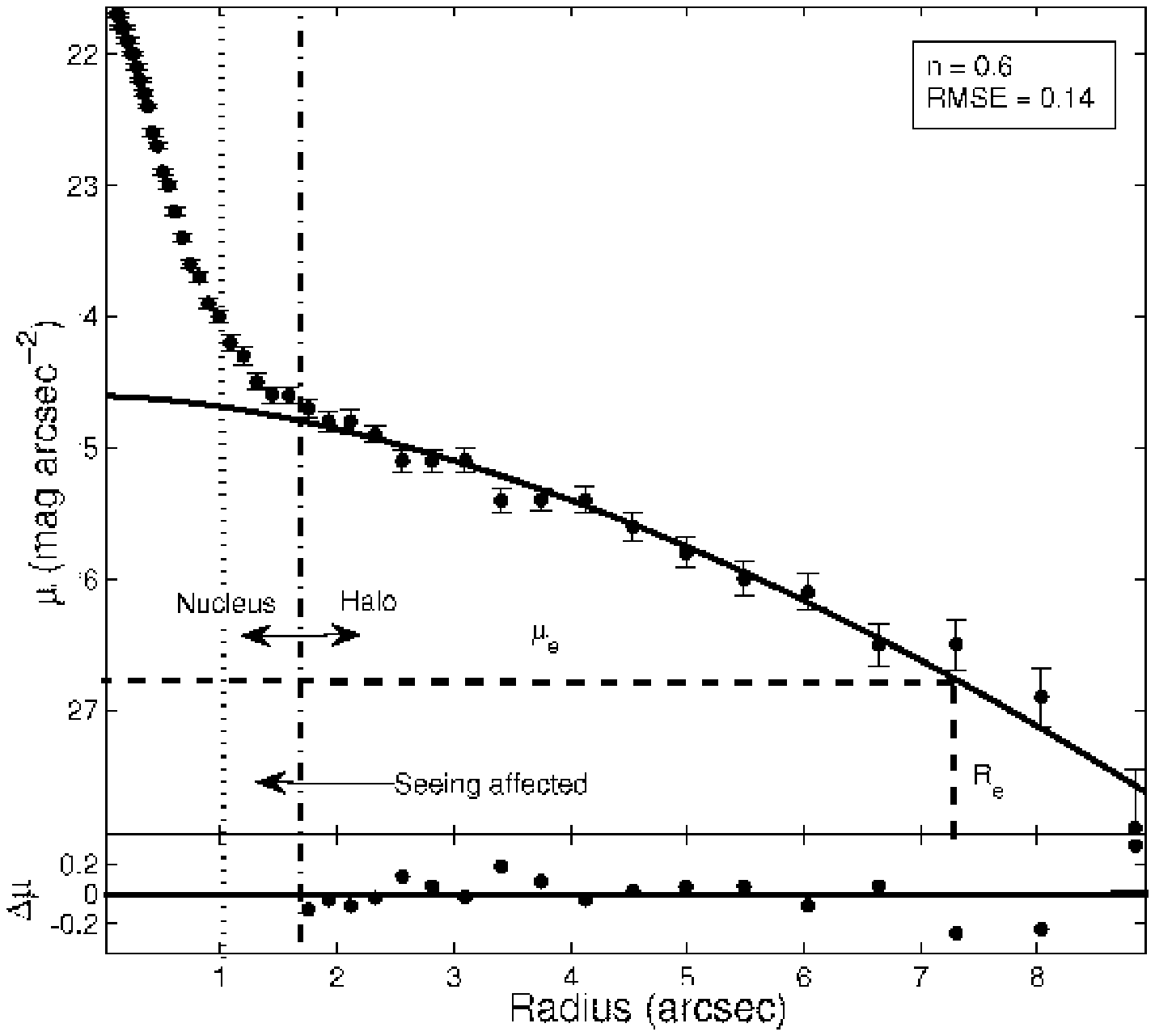,width=9cm}
  \epsfig{figure=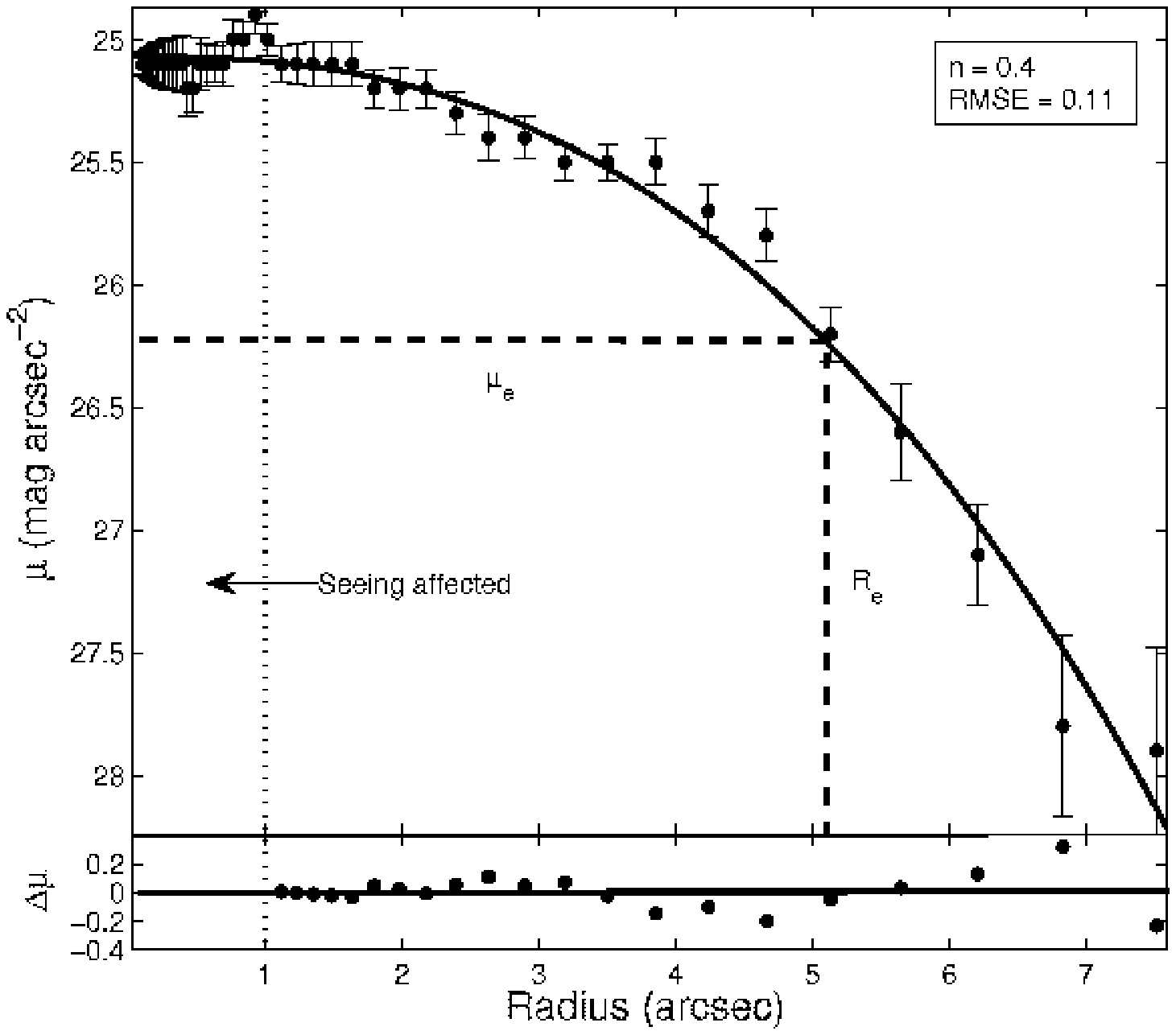,width=9cm}
  \caption{\textit{Left:} Knot A's halo surface brightness profile obtained with IRAF's
\textit{ellipse} task. The continuos line represent the S\'ersic n=0.6
model fit. The dotted line delimits the seeing affected zone, the
dash-dotted line have been placed at $R_{core}=1.6\arcsec$. 
Since the nucleus is only partially resolved,
data points below $R_{core}$ have been excluded from the fit. \textit{Right:} Same as previous for knot B. In this case the S\'ersic index is n=0.4 and data points below $1\arcsec$ have been excluded from the fit.}
\label{fig4}
\end{center}
\end{figure*}

\begin{figure*}[h!]
\begin{center}
  \epsfig{figure=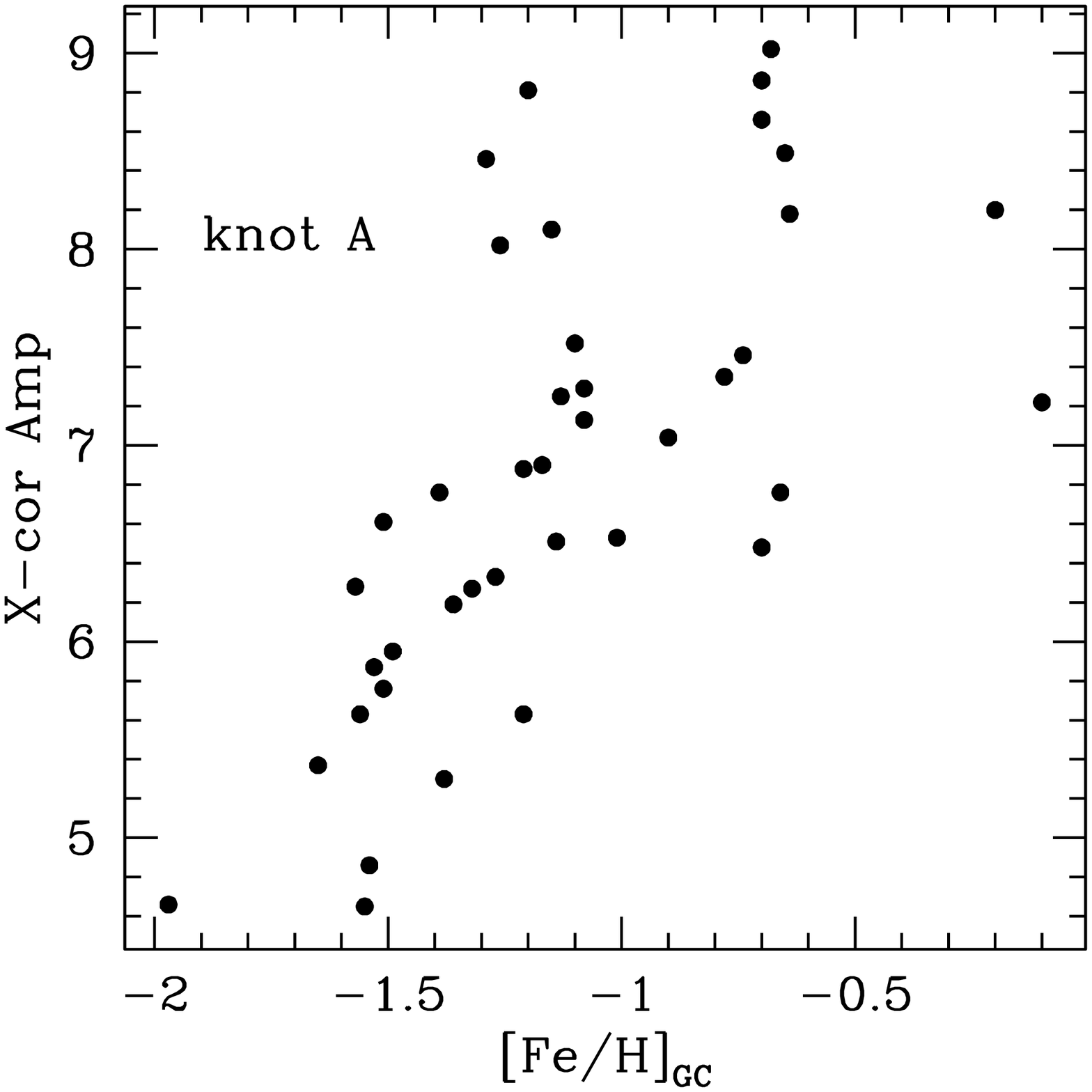, width=5 cm}
  \epsfig{figure=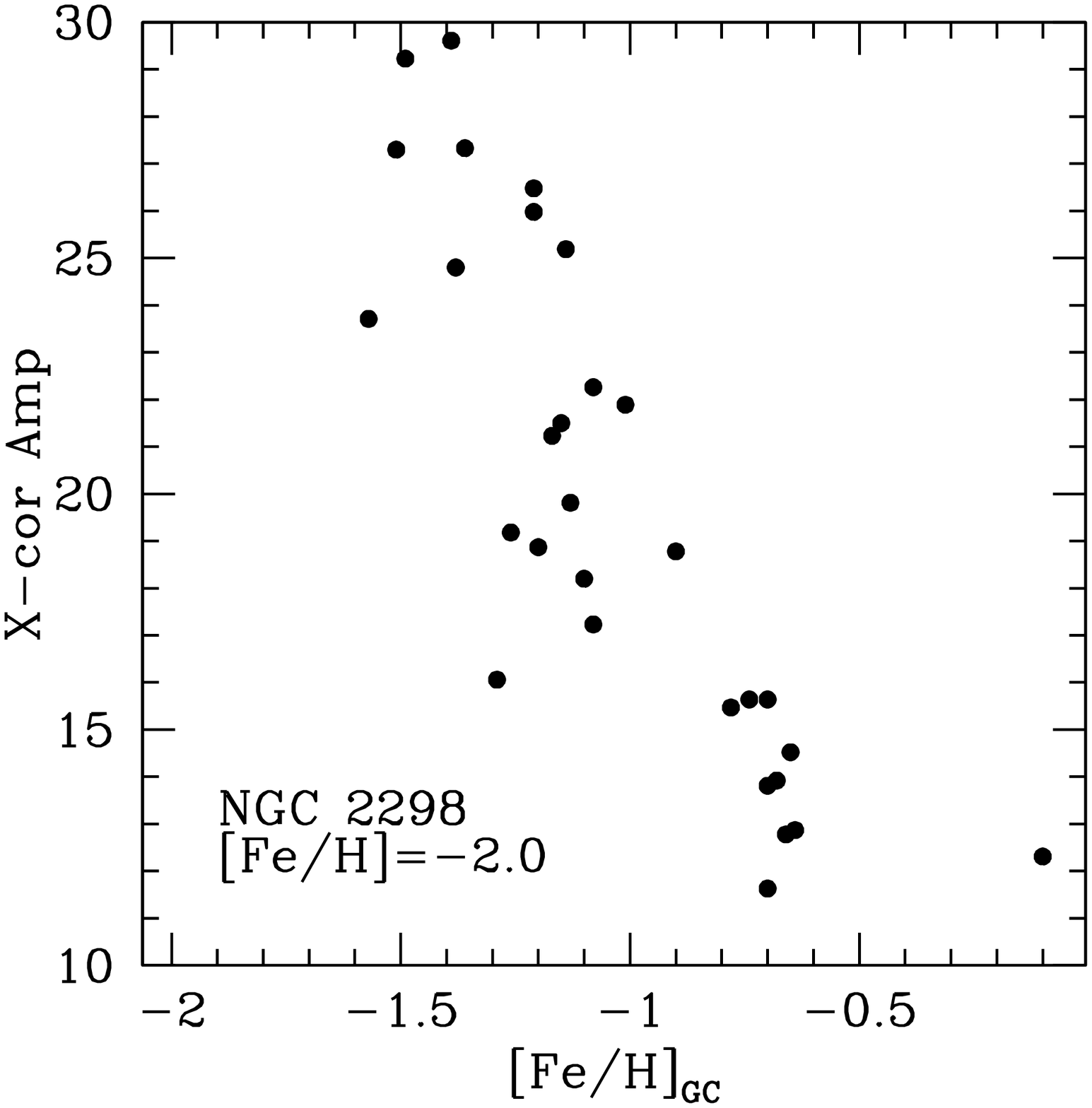, width=5 cm}
  \epsfig{figure=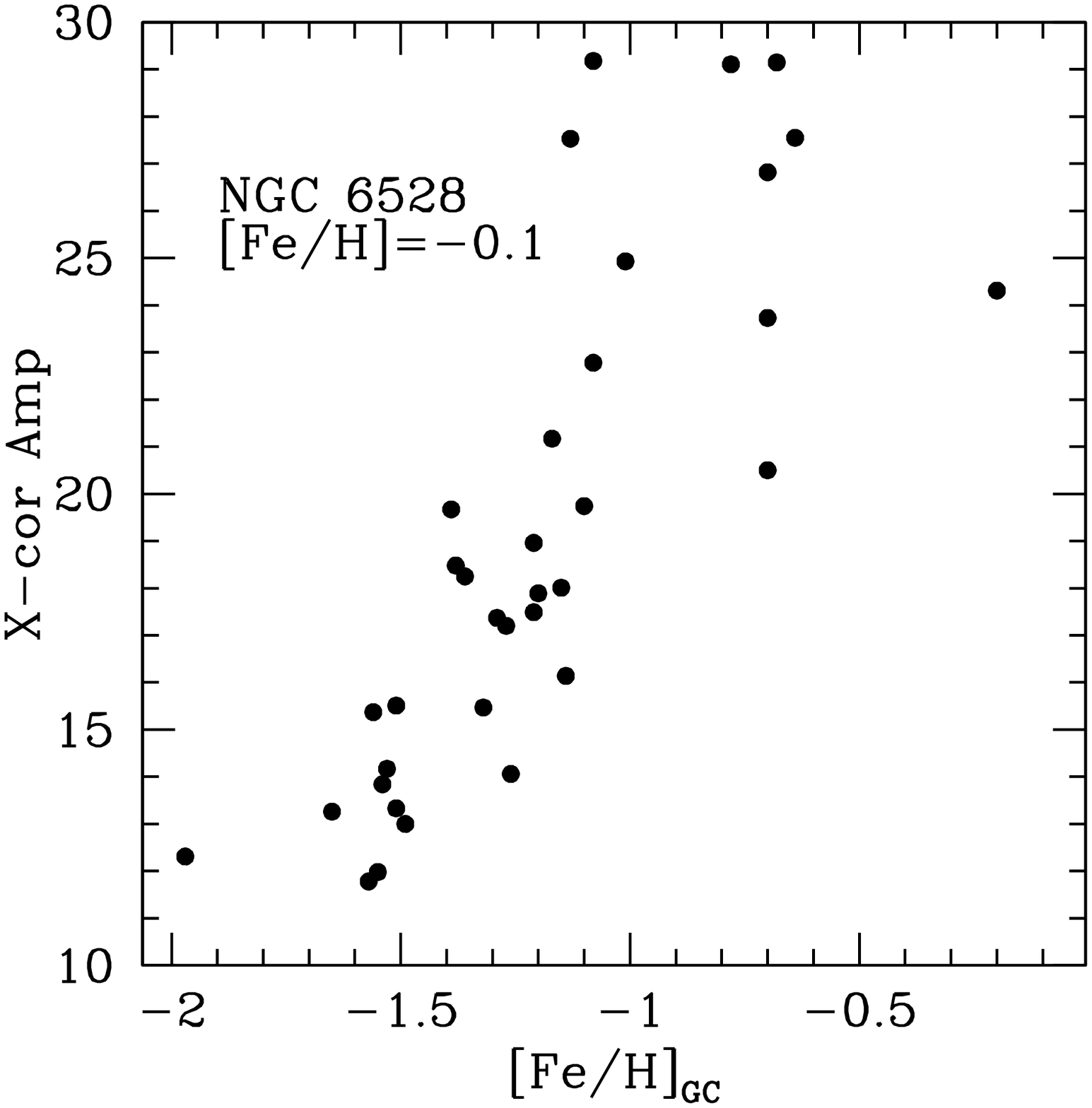, width=5 cm}
  \caption{\textit{Left:} Knot A's cross correlation amplitude as a function of 
  metallicity for galactic GCs from Schiavon (2005). The rough linear relation of positive 
  slope, indicates that our spectrum is better fit by metal-rich than metal-poor GCs 
 (see the text and next figure for details). \textit{Center, Right:} The same as left for 
 two Galactic GCs from the same library, whose metallicity has been estimated with HR spectroscopy. 
 As expected, the metal poor GC NGC 2298 correlates better with metal poor GCs while the metal rich GC
 NGC 6528 correlates best with metal rich GCs.}
\label{fig5}
\end{center}
\end{figure*}

\begin{figure*}[h!]
\begin{center}
  \epsfig{figure=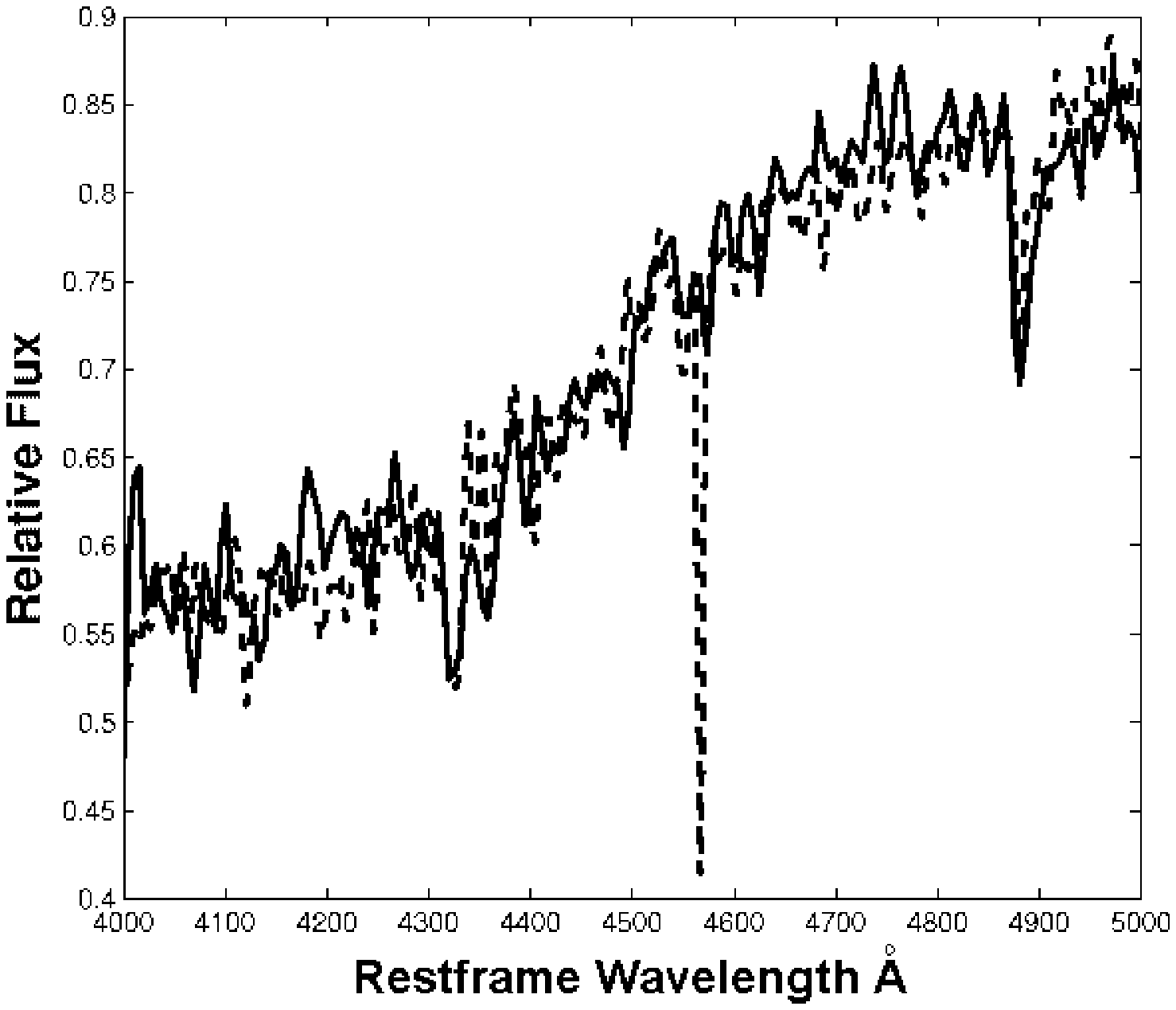,width=9cm}
  \epsfig{figure=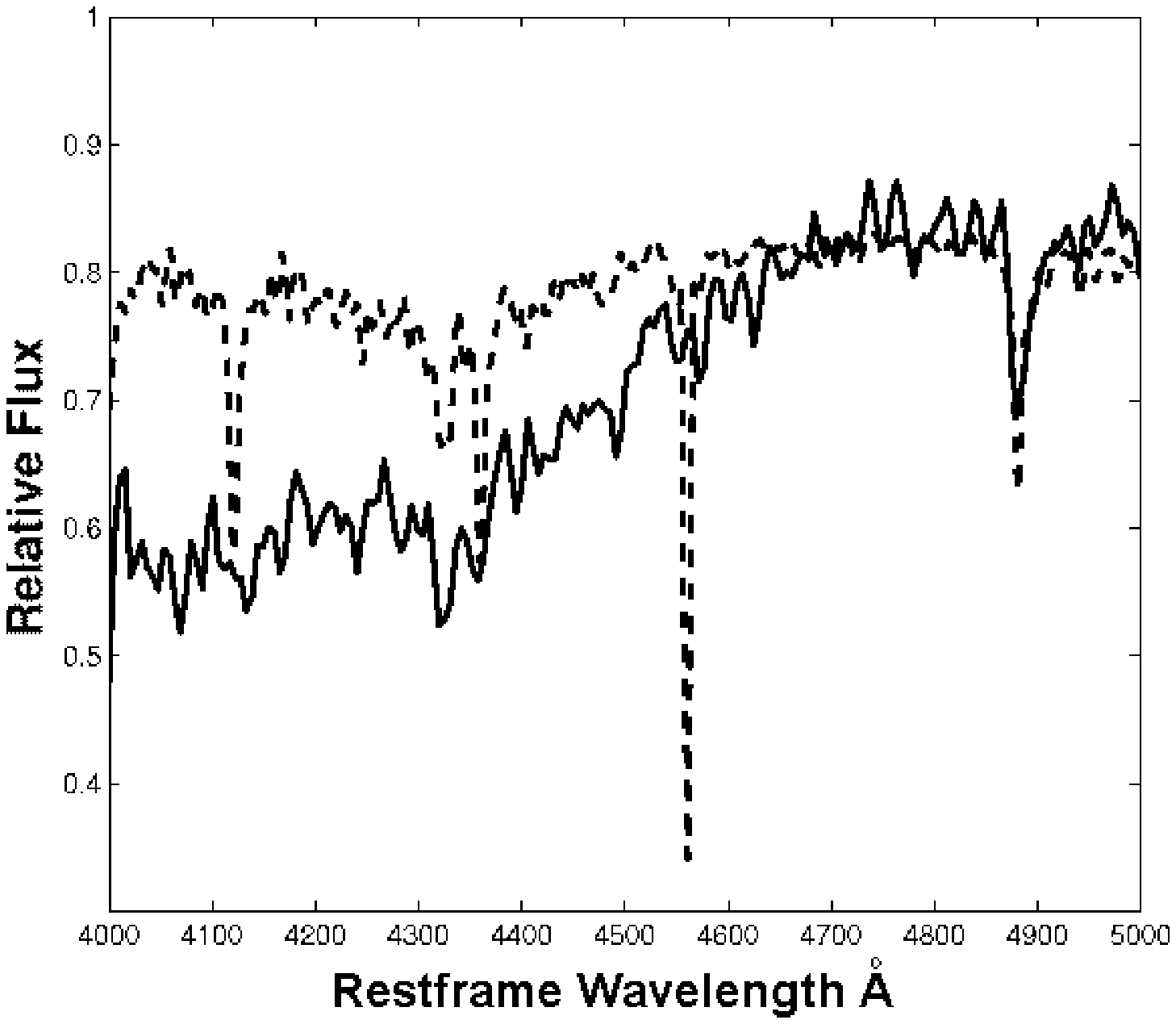,width=9cm}
  \caption{\textit{Left:} Comparison between knot A's spectrum (continuous line) and the best 
  correlating GC from Schiavon (2005) (dashed line, cross correlation amplitude 9) which is the intermediate
  metallicity GC NGC~6388. \textit{Right:} The same as left with the worst correlating GC from
  Schiavon et al. (2005) (cross correlation amplitude 4.5), which is NGC~1904. The two plots shows that knot A's
  spectrum correlates best with intermediate to high metallicity GCs like NGC~6388 ($[Fe/H] = -0.7$), than 
  with metal poor ones like NGC~1904 ($[Fe/H] = -1.5$).}
\label{fig6}
\end{center}
\end{figure*}

\begin{figure*}
\resizebox{\hsize}{!}{\includegraphics{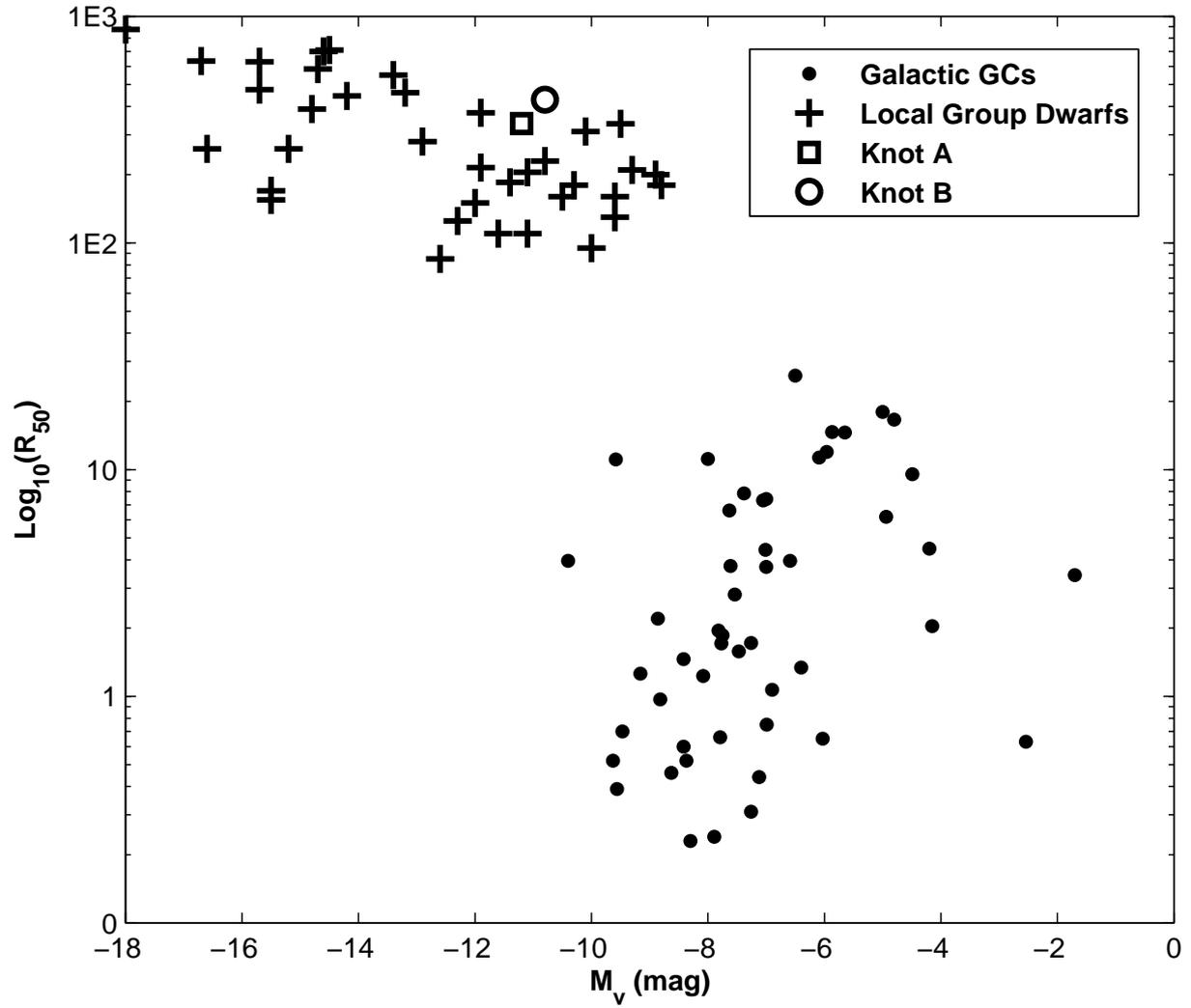}}
\caption{Absolute magnitudes vs half light radii for local group dwarfs from (\cite{mat98}), Galactic globular clusters (\cite{web85}) and the two knots.}
\label{fig7}
\end{figure*}

\begin{figure*}[h!]
\begin{center}
  \epsfig{figure=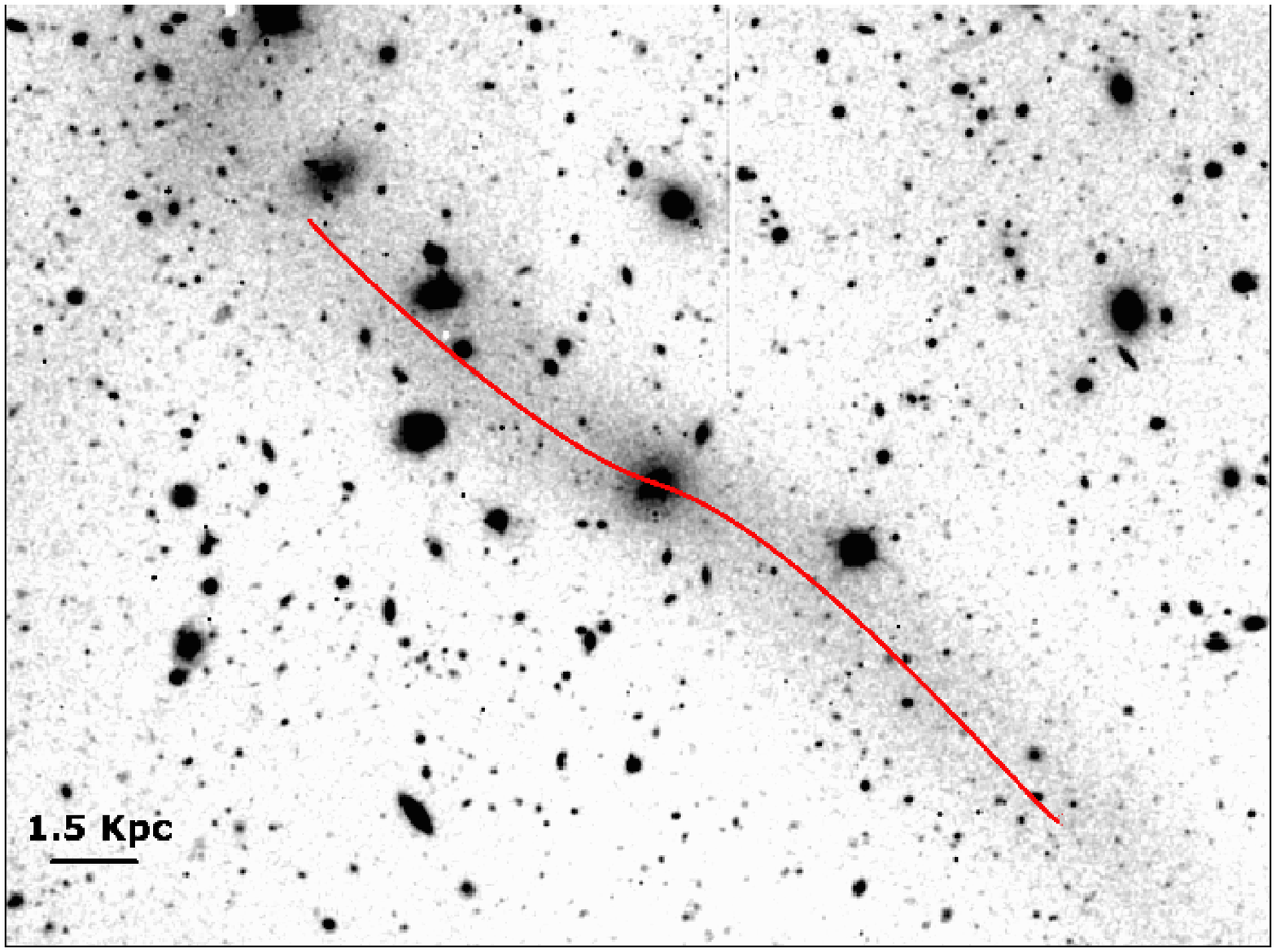,width=9cm}
  \epsfig{figure=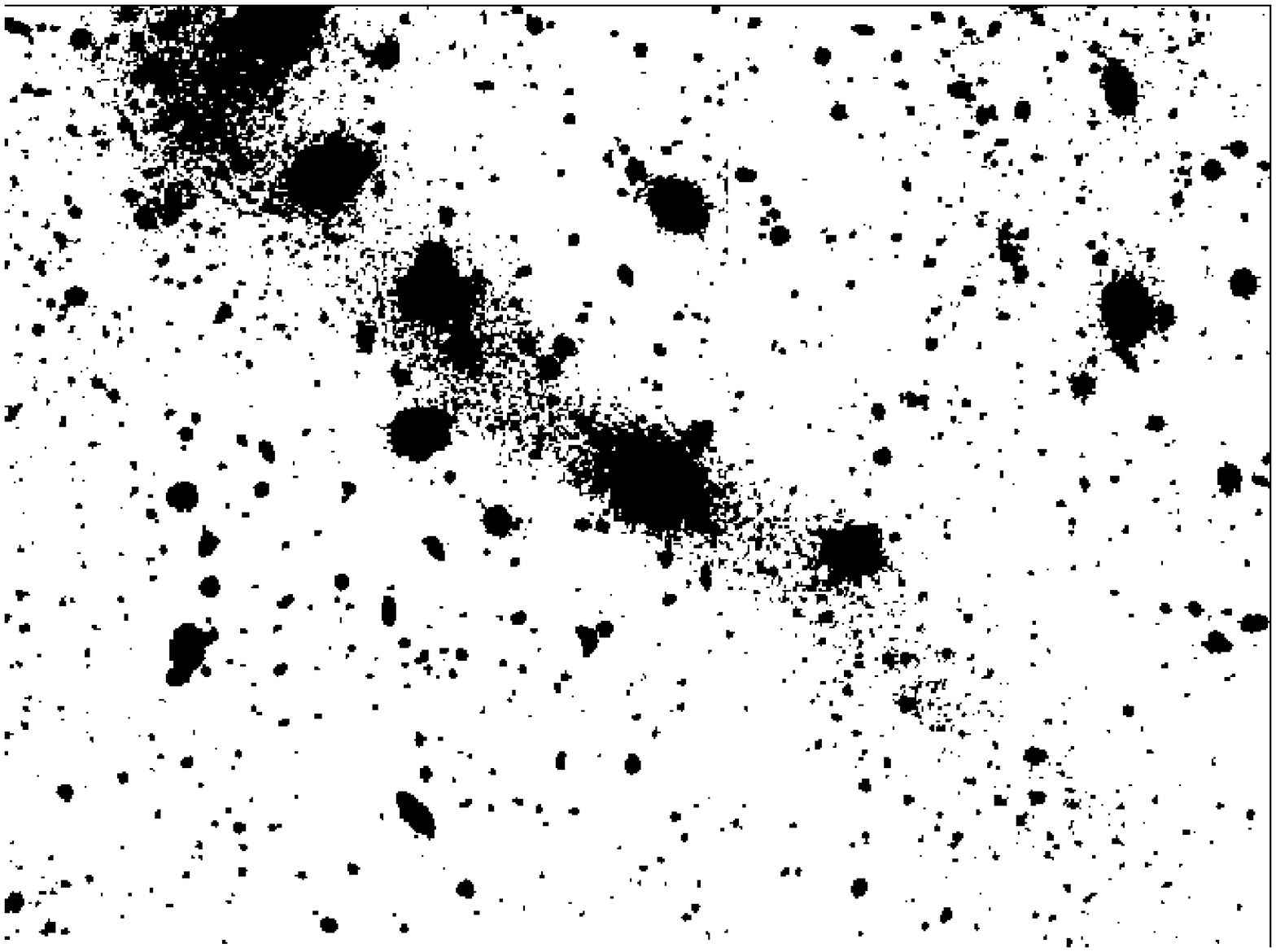,width=9cm}
  \caption{\textit{Left:} In this ehnanced version of Fig. 1 (3\arcsec gaussian smoothing, logarithmic scaling of intensities)
  we show how the stream forms an ''S" shaped inflection in correspondence of knot A. \textit{Right:} In this contrast-ehnanced
  elaburation of Fig 1. the elliptical overdensity of the stream at the projected position of knot A is evident. Both features
  are typical for tidally disrupting systems.}
\label{fig8}

\end{center}
\end{figure*}


\begin{thebibliography}{}
\bibitem[Arp 1976]{arp76} Arp, H. 1976, \apj, 207L,147A
\bibitem[Appenzeller et al. 1998]{app98} Appenzeller, I. et al. 1998, The Messenger, 94,1
\bibitem[Bekki et al. 2003]{bek03} Bekki, K. et al. 2003, \mnras, 344, 399
\bibitem[Bekki, K. et al. 2003]{bek032} Bekki, K. et al., 2003, \mnras, 344, 1334
\bibitem[Bell \& de Jong 2001]{bel01} Bell, E. F. \& de Jong, R. S. 2001, \apj, 550, 212
\bibitem[Bruzual \& Charlot (2003)]{bc03} Bruzual, G. \& Charlot, S. 2003, \mnras, 344, 1000B
\bibitem[Carter, Allen \& Malin 1984]{car84} Carter, D., Allen, D. A., Malin, D. F. 1984, \mnras, 211, 707
\bibitem[Capaccioli 1987]{cap87} Capaccioli, M. 1987 in IAU Symp. 127, Structure and Dynamics of Elliptical Galaxies, ed. P. T. de Zeeuw (Dordrecht: Reidel), 47
\bibitem[Dekel \& Silk 1986]{dek86} Dekel, A., Silk, J. 1986 \apj, 303, 39
\bibitem[D'Onghia et al. 2009]{don09} D'Onghia, E. et al. 2009, \nat, 460, 605D 
\bibitem[D'Onghia \& Lake 2008]{don08} D'Onghia, E., Lake, G. 2008, \apj, 686L, 61D
\bibitem[Duc \& Mirabel 1994]{duc94} Duc, P. A., Mirabel, I. F. 1994, \aap, 289, 83
\bibitem[de Vancouleurs 1948]{van48} de Vancouleurs, G 1948, Ann. d'Astrophysique, 11, 247
\bibitem[de Vancouleurs 1959]{van59} de Vaucouleurs, G. 1959, Handb. Phys., 53, 275
\bibitem[Drinkwater et al. 2000]{dri00} Drinkwater, M. J. et al. 2000, \pasa, 17, 227
\bibitem[Faber \& Lin 1983]{feb83} Faber, S. M., Lin D. N. C., \apj 266L, 17F  
\bibitem[Fellhauer \& Kroupa 2002]{fel04} Fellhauer, M., Kroupa, P. 2002, \mnras, 330, 642
\bibitem[Fellhauer \& Kroupa 2005]{fel05} Fellhauer, M., Kroupa, P. 2005, \mnras, 359, 223
\bibitem[Forbes et al. 2003]{for03} Forbes, D. et al., 2003, Science, 301, 1217F
\bibitem[Genzel et al. 2006]{gen06} Genzel, R. et al. 2006, \nat, 442, 17
\bibitem[Graham 2001]{gra01} Graham, A. W., 2001, \aj, 121, 820
\bibitem[Graham 2003]{gr03} Graham, A. W. \& Guzman, R. 2003, \apj, 125, 2936
\bibitem[Gratton et. al. 2003]{gra03} Gratton, R. G. et al. 2003, \aap, 408, 592
\bibitem[Grebel, Gallagher \& Harbeck 2003]{gre03} Grebel, Eva K., Gallagher, John S., III; Harbeck, Daniel, 2003, \aj, 125, 1926
\bibitem[Harris 2005]{harr05} Harris, W. E., 1996, \aj, 112, 1487
\bibitem[Hasegan et al. 2005]{hase05} Hasegan, M. et al. 2005, \apj, 627, 203
\bibitem[Hibbard et al. 1994]{hib94} Hibbard, J. E et al. 1994, \aj, 107, 67
\bibitem[Higdon \& Wallin 2003]{hig03} Higdon, J. L., Wallin, J. F. 2003, \apj, 585, 281
\bibitem[Higdon, Higdon \& Marshall  2006]{hig06} Higdon, S. J., Higdon, J. L., Marshall, J. 2006, \apj, 640, 768
\bibitem[Hilker, Infante \& Richtler 1999]{hil991} Hilker, M., Infante, L., Richtler, T. 1999, \aap, 138, 55
\bibitem[Hilker et al. 1999]{hil992} Hilker, M. et al. 1999, \aap, 134, 75
\bibitem[Hilker et al. 1999]{hil993} Hilker, M. et al. 1999, \aap, 138, 59
\bibitem[Kissler-Patig, Jordan \& Bastian 2006]{kis06} Kissler-Patig, M., Jordan, A., Bastian, N., 2006, \aap, 448, 1031
\bibitem[Koribalski et al. 2004]{kori04} Koribalski, B. et al. 2004, \aj, 128, 16
\bibitem[Kroupa et. al 2010]{kro10} Kroupa, P. et al. 2010, \aap, in press
\bibitem[Kroupa 1997]{kro97} Kroupa, P. et al. 1997, New Astronomy 2, 139, 164
\bibitem[Kroupa, Theis \& Boily 2005]{kro05} Kroupa, P., Theis, C., Boily C.M., 2005, \aap 2, 431, 517K
\bibitem[Lasker et al. 1990]{las90} Lasker, B., M. et al. 1990, \aj, 99, 2019
\bibitem[Lorre 1978]{lor78} Lorre, J., J. 1978, \apj, 222, 99
\bibitem[Li \& Helmi 2008]{li08} Li, Y., Helmi, A. 2008 \mnras, 385, 1365
\bibitem[Lianou, Grebel \& Kock 2010]{li10} Lianou, S., Grebel, E. K., Koch A. 2010 \aap, in press
\bibitem[Lynden-Bell \& Lynden-Bell 1995]{lili95} Lynden-Bell, D., Lynden-Bell, R. M. 1995 \mnras, 275, 429
\bibitem[Mackey \& van den Bergh]{macksidney} Mackey, A. D., \& van den Bergh, S. 2005, \mnras, 360, 631
\bibitem[Maraston et al. 2004]{mar04} Maraston, C., et al. 2004, \aap, 416, 467
\bibitem[Mart\'inez-Delgado et al. 2010]{mar10} Mart\'inez-Delgado D. et al., 2010, submitted to ApJL, 2010arXiv1003.4860M
\bibitem[Mart\'inez-Delgado et al. 2008]{mar08} Mart\'inez-Delgado D. et al., 2008, \apj, 689, 184M
\bibitem[Mateo 1998]{mat98} Mateo, M. 1998, \araa, 435, 506
\bibitem[Mathewson \& Ford]{mf96} Mathewson, D. \& Ford, V. 1996, \apjs, 107, 97
\bibitem[Metz \& Kroupa 2007]{mk07} Metz, M., Kroupa, P. 2007 \mnras, 376, 387
\bibitem[Metz, Kroupa and Jerjen 2009]{mk09} Metz, M., Kroupa, P., Jerjen, H. 2009 \mnras, 394, 2223
\bibitem[Metz et al. 2009]{mz09} Metz, M. et al. 2009 \apj, 697, 269
\bibitem[Meylan et al. 2001]{mey01} Meylan, G. et al. 2001, \apj, 122, 830
\bibitem[Meylan et al. 1995]{mey95} Meylan, G., Mayor, M., Duquennoy, A., \& Dubath, P. 1995, \aap, 303, 761
\bibitem[Mieske et al. 2006]{mie06} Mieske, S. et al. 2006, \apj, 131, 2242
\bibitem[Mirabel, Dottori \& Lutz 1992]{mir92} Mirabel, I. F., Dottori, H., Lutz D. 1992, \aap, 256L, 19M
\bibitem[Misgeld, Hilker \& Mieske 2008]{mhm08} Misgeld, I., Hilker, M., Mieske, S. 2008 \aap, 486, 697M
\bibitem[Misgeld, Hilker \& Mieske 2009]{mhm09} Misgeld, I., Hilker, M., Mieske, S. 2009 \aap, 496, 683M
\bibitem[Monet et al. 2003]{mon03} Monet, D. G., et al. 2003, \aj, 125, 984
\bibitem[Meylan \& Mayor 1986]{mey86} Meylan, G. and Mayor, M. 1986, \aap, 166, 122
\bibitem[Okazaki \& Taniguchi 2000]{oka00} Okazaki, T., Taniguchi, Y. 2000 \apj, 543, 149
\bibitem[Patat, F. 2003]{pat03} Patat, F., 2003, ''From twilight to highlight: the physics of Supernovae", Hillebrandt, W. and Leibundgut, B. eds., Springer, 321
\bibitem[Pe\~narrubia et al. 2009]{pen09} Pe\~narrubia, J., et al. 2009, \apj, 698, 222
\bibitem[Pe\~narrubia, McConnachie \& Navarro 2008]{pen08} Pe\~narrubia, J., McConnachie A. W., Navarro, J. F. 2008, \apj, 673, 226P
\bibitem[Phillips at al. 2001]{phi01} Phillips, S. et al. 2001, \apj, 560, 201
\bibitem[Prieto et al. 2005]{pri05} Prieto, M. et al. 2005, \aj, 130, 1472
\bibitem[Ricotti et. al. 2008]{ric08} Ricotti, M. et al. 2008, \apj, 685, 21
\bibitem[Scarpa 2006]{sca06} Scarpa, R. 2006 (preprint: astro-ph/0504051) 
\bibitem[Schweizer 1982]{sch82} Schweizer, F. 1982, \apj,  252, 455
\bibitem[Schiavon et al. 2005]{schi05} Schiavon, R. P., et al. 2005, \apjs, 160, 163
\bibitem[S\`ersic 1968]{ser68} S\`ersic, J. L. 1968, Atlas de galaxias australes (Cordoba, Argentina: Observatorio Astronomico, 1968)
\bibitem[Spergel et al. 2003]{spe03} Spergel, N.D. et al. 2003, \apjs, 148, 175-194
\bibitem[Storchi-Bergmann et al. 2005]{sto05} Storchi-Bergmann, T. et al. 2005, \apj, 624, 13S
\bibitem[Webbink 1985] {web85} Webbink, R.F. 1985 IAU Symp 113 ''Dynamics of Star Clusters", Ed. J. Goedman \& P. Hut, 541 (1985)
\bibitem[Wehrle et al. 1997]{weh97} Wehrle, A. E., et al. 1997, \aj, 114, 115
\bibitem[Wolstencroft et al. 1984]{wol84} Wolstencroft, R. D., Tully, R. B., Perley, R. A. 1984, \mnras, 207, 889
\bibitem[Knierman et al. 2003]{kni03} Knierman, K. A. et al. 2003, \aj, 126, 1227
\bibitem[Vazdekis 1999]{vaz99} Vazdekis, A. 1999, \apj, 513, 224
\bibitem[Zwicky 1956]{zwi56} Zwicky, F. 1956, Ergenisse Exakten Naturwissenschaften, 29, 344
\end{thebibliography}
\end{document}